\documentclass[%
reprint,
superscriptaddress,
 amsmath,amssymb,
 aps,
]{revtex4-2}

\usepackage{graphicx}
\usepackage{dcolumn}
\usepackage{bm}
\usepackage[colorlinks=true,linkcolor=black]{hyperref}

\usepackage{array}
\usepackage{physics}
\renewcommand\vec{\mathbf} 
\usepackage{xr}
\makeatletter
\newcommand*{\addFileDependency}[1]{
\typeout{(#1)}
\@addtofilelist{#1}
\IfFileExists{#1}{}{\typeout{No file #1.}}
}\makeatother
\newcommand*{\myexternaldocument}[1]{%
\externaldocument{#1}%
\addFileDependency{#1.tex}%
\addFileDependency{#1.aux}%
}
\myexternaldocument{SI}

\makeatletter
\AtBeginDocument{\@ifpackageloaded{natbib}{\ifNAT@numbers\if@filesw\immediate\write\@auxout{\string\global\string\NAT@numberstrue}\fi\fi}{}

\let\($\let\)$}
\makeatother

\usepackage{amstext}

\begin{document}

\preprint{APS/123-QED}

\title{Statistical modeling of equilibrium phase transition in confined fluids}

\author{Gunjan Auti} 
\email{gunjanauti@thml.t.u-tokyo.ac.jp}
\affiliation{
Department of Mechanical Engineering, The University of Tokyo, \\7-3-1 Hongo, Bunkyo-ku, Tokyo 113-8656, Japan}

\author{Soumyadeep Paul}
\affiliation{
Department of Mechanical Engineering, The University of Tokyo, \\7-3-1 Hongo, Bunkyo-ku, Tokyo 113-8656, Japan}
\affiliation{
Department of Mechanical Engineering, Stanford University, Building 530, 440 Escondido Mall, Stanford, California 94305, USA}

\author{Wei-Lun Hsu}
\affiliation{
Department of Mechanical Engineering, The University of Tokyo, \\7-3-1 Hongo, Bunkyo-ku, Tokyo 113-8656, Japan}

\author{Shohei Chiashi}
\affiliation{
Department of Mechanical Engineering, The University of Tokyo, \\7-3-1 Hongo, Bunkyo-ku, Tokyo 113-8656, Japan}

\author{Shigeo Maruyama}
\affiliation{
Department of Mechanical Engineering, The University of Tokyo, \\7-3-1 Hongo, Bunkyo-ku, Tokyo 113-8656, Japan}

\author{Hirofumi Daiguji}
\email{daiguji@thml.t.u-tokyo.ac.jp}
\affiliation{
Department of Mechanical Engineering, The University of Tokyo, \\7-3-1 Hongo, Bunkyo-ku, Tokyo 113-8656, Japan}%

\date{\today}

\begin{abstract}
The phase transition of confined fluids in mesoporous materials deviates from that of bulk fluids due to the interactions with the surrounding heterogeneous structure. For example, adsorbed fluids in metal-organic-frameworks (MOFs) have atypical phase characteristics such as capillary condensation and higher-order phase transitions due to a strong heterogeneous field. Considering a many-body problem in the presence of a nonuniform external field, we model the host-guest and guest-guest interactions in MOFs. To  solve the three-dimensional Ising model, we use the mean-field theory to approximate the guest-guest interactions and Mayer's \(f\)-functions to describe the host-guest interactions in a  unit cell. Later, using Hill's theory of nanothermodynamics, we define differential thermodynamic functions to understand the distribution of intensive properties  and integral thermodynamic functions to explain the phase transition in confined fluids. The investigation reveals a distinct behavior where fluids confined in larger pores  undergo a discontinuous (first-order) phase transition, whereas those confined in smaller pores experience a continuous (higher-order) phase transition. Furthermore, the results indicate that the free-energy barrier for phase transitions is lower in confined fluids than in  bulk fluids giving rise to a lower condensation pressure relative to the bulk saturation pressure. Finally, the integral thermodynamic functions are succinctly presented in the form of a phase diagram, marking an initial step toward a more practical approach for understanding the phase behavior of confined fluids.
\end{abstract}

\maketitle


\section{Introduction}
\label{sec:introduction} 
Confined fluids can be gas (vapor) trapped in nanobubbles \cite{taverna2008probing}, adsorbed gas in porous structures \cite{sircar1970statistical,evans1990fluids,morris2008gas}, natural gas trapped in shale and tight rock formations \cite{jew2022chemical}, or biomolecules trapped in  cells \cite{krishnamurthy2012engineering}. Fluids such as these enclosed within restricted spaces have distinct physical and thermodynamic characteristics that differ from those of bulk fluids. These unique characteristics include phenomena such as an atypical phase transition and packing polymorphism  \cite{alabarse2012freezing,algara2015square,gimondi2018co,kapil2022first}, a shift in the freezing and melting points \cite{hamada2007phase}, an anomalously low dielectric constant \cite{fumagalli2018anomalously}, and very high hydrodynamic slippage \cite{kavokine2022fluctuation}. Although these phenomena have been widely observed and reported, how heterogeneity and their multiscale nature affect the fluid characteristics still lacks a comprehensive thermodynamic understanding. 

The atypical thermodynamic properties are due to the heterogeneous interactions and steric hindrance of  confined fluids. Generally, these interactions take the form of van der Waals forces and the inverse radial dependence of these cohesive interactions translates into a layered distribution of density near the surface, which creates anisotropy \cite{de1981polymer, fukui1999imaging, fukuma2010atomic, page20143, wang2018layered}. Gibbs laid the foundation for modeling how heterogeneous interactions affect fluid properties by  formulating surface thermodynamics, where he introduced the concept of  the \textit{Gibbs surface excess} \cite{gibbs1928collected}. Later, Hill put forth a thermodynamic approach to model  small systems \cite{hill1962thermodynamics}. Since then, numerous models   describing how heterogeneity affects the fluid properties have been proposed. For example, various self-consistent field models have been presented based on the analogous Bogoliubov--Born--Green--Kirkwood--Yvon  hierarchy describing a system containing a large number of interacting particles \cite{green1960molecular, steele1960distribution, wertheim1984fluids, wertheim1984fluids2, wertheim1986fluids3, wertheim1986fluids4, biesheuvel2006self, poluektov2015thermodynamic, laktionov2022colloidal}. These models can be solved for special cases but become much more complex for a different sets of conditions.

To make the model more realizable under general conditions, Sircar and Mayers \cite{sircar1970statistical, sircar1985excess} proposed a semi-empirical formulation for low-concentration adsorption. They coined the term ``ideal adsorbed phase" to indicate a behavior that differs from that of  bulk fluids. Nicholson \cite{nicholson1975molecular,nicholson1976molecular} later presented an elaborate molecular theory focusing on the adsorption of lattice gas. This model shows qualitatively  how the  thermodynamic properties vary but  does not focus on the phase transition of the adsorbed fluid. Martinez \textit{et al.} \cite{martinez2007predicting} predicted adsorption isotherms using a two-dimensional statistical associating fluid theory  for a square well potential on a flat surface. 

In parallel, with the advancement of molecular simulations, Evans \cite{evans1986fluids, evans1990fluids} undertook extensive Monte Carlo  simulations to elucidate phase transitions in mesoporous slits. Schmidt and L{\"o}wen \cite{schmidt1997phase} introduced a hard-sphere model to investigate the freezing transition between parallel plates, presenting a phase diagram through Monte Carlo simulations. Subsequently, numerous studies used brute force molecular simulations to produce both qualitative and quantitative phase diagrams. For instance, Kimura and Maruyama \cite{kimura2002molecular} explored boiling phase transitions and cluster formation using molecular dynamics  simulations, while Radhakrishnan \textit{et al.} \cite{radhakrishnan2002global} used biased potentials and umbrella sampling in grand canonical Monte Carlo (GCMC) simulations to obtain thermodynamic properties. 
Takaiwa \textit{et al.} \cite{takaiwa2008phase} presented a phase diagram of water in carbon nanotubes. Zhou \textit{et al.} \cite{zhou2019determining} conducted \textit{ab initio} simulations using density functional theory  and grand canonical algorithms to illustrate a surface phase diagram. 

More recently,  these molecular simulation techniques have been improved by incorporating various machine-learning algorithms \cite{zhang2019prediction, gurnani2021interpretable, fung2021machine, cui2023direct}. Such methodologies aim to extract specific properties of adsorbed fluids, reducing computational demands and facilitating the prediction of certain thermodynamic properties. However, even the most current  deep-learning models are black boxes \cite{castelvecchi2016can,rudin2019stop}, meaning that they perform the majority of their calculations internally, so that significant thermodynamic information is overlooked. This hinders a comprehensive understanding of the physics underlying adsorption and confinement. Moreover, deep-learning algorithms are essentially the methods of statistical mechanics, as explained by Lin \textit{et al.} \cite{lin2017does}.
Consequently, understanding the thermodynamics of confined fluids requires  statistical methods and analytical models based on appropriate approximations. 

Statistical models have been used to understand various multiscale processes. For example, Ko\v{s}mrlj and Nelson \cite{PhysRevX.7.011002} gave a model for the thin shells and argued that large spherical shells are unstable due to thermally generated pressure using statistical mechanics. Goodrich \textit{et al.} \cite{goodrich2021designing} formulated a statistical model for nanocluster formation in the crystallization process, and Molina \textit{et al.} \cite{molina2021droplet} experimentally described the many-body interactions that occur in confined space for self-organizing of droplets. In a similar way, to understand the multiscale process of phase transition in confined fluids, we have developed a semi-analytical statistical model.

\begin{figure}
    \centering
    \includegraphics{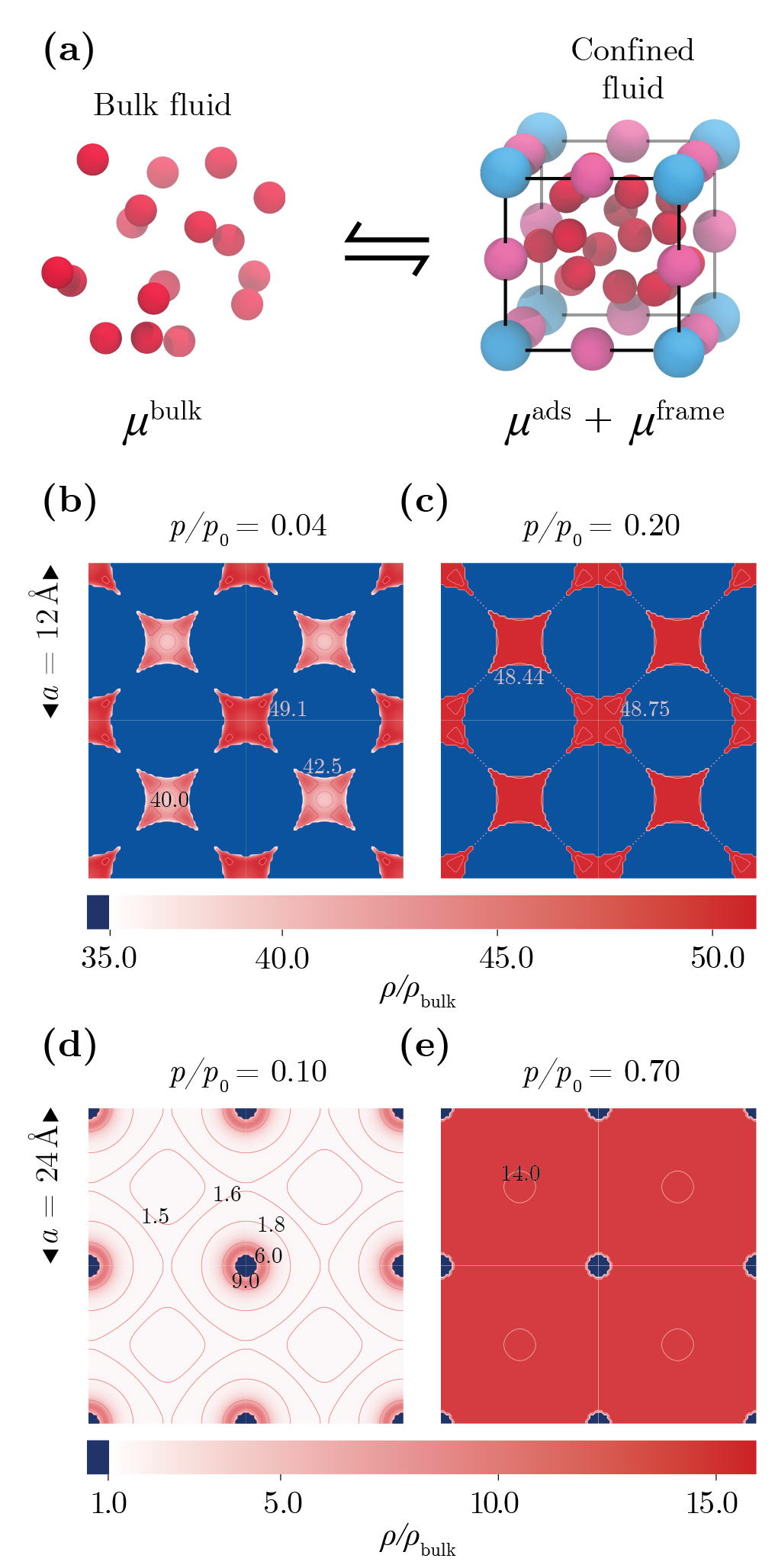}
    \caption{Statistical model for adsorbed fluid.
    (a) Schematic representation of the system under consideration. The argon molecules (red) in bulk with  chemical potential \(\mu^{\text{bulk}}\) are in equilibrium with the system of adsorbed molecules with chemical potential \(\mu^{\text{ads}}\) in a metal-organic framework (MOF) with chemical potential \(\mu^{\text{frame}}\). The relative density distribution obtained from the proposed model for an MOF with a 12 \AA{} unit cell is shown at (b) low relative pressure (\(p/p_0 = 0.04\)) and (c) after saturation (\(p/p_0 = 0.20\)). Similar distribution functions are plotted for a 24 \AA{} unit cell MOF at (d) low relative pressure (\(p/p_0 = 0.10\)) and (e) after saturation (\(p/p_0 = 0.70\)). The deep blue shade represents the heterogeneity (metal or ligand) and all  distributions are plotted for a cross section of the unit cell at \(z=a/2\).   }
    \label{fig:figure1}
\end{figure}

The present paper introduces a three-dimensional (3D) Ising model for argon confined in a cubic metal-organic framework [MOF, see Fig.~\ref{fig:figure1}(a)]. This approach considers the nonuniformity of the external field by decoupling homogeneous and heterogeneous interactions. The homogeneous interactions are considered through mean-field theory and the heterogeneous interactions are approximated using Mayer's \(f\)-functions. The nonuniform interactions lead to a nonuniform density distribution in the pores, as depicted in  Figs. \ref{fig:figure1}(b)--\ref{fig:figure1}(e). To account for the thermodynamic properties of such a distribution, differential (local) and integral (global) intensive thermodynamic functions can be defined.  For example, Figs. \ref{fig:figure1}(b) and  \ref{fig:figure1}(c) show the expected relative density distribution (a differential property) of argon at relative pressures, \(p/p_0 = 0.04\) and \(p/p_0 = 0.20\), respectively, for a small pore (\(a = 12\) \AA). Even for extremely low relative pressure, the pores are close to saturation. Conversely, Figs. \ref{fig:figure1}(d) and  \ref{fig:figure1}(e) show the expected relative density distribution of argon at relative pressures \(p/p_0 = 0.10\) and \(p/p_0 = 0.70\), respectively, for a large pore (\(a = 24\) \AA). Here, the layered adsorption occurs near the heterogeneity at lower relative pressure and a uniform density distribution develops at higher relative pressure. Later in this paper, we  derive the integral thermodynamic functions using Hill's thermodynamics for small systems. Based on these integral properties, we discuss the phase transition of confined fluids and compare it with that of bulk fluids. 

This paper discusses the statistical modeling of the confined fluid while highlighting the phase transition during capillary condensation. We also showcase the phase diagram of the adsorbed fluid and discuss the similarities and key differences vis-\'a-vis  the bulk fluid. The remainder of the paper is organized as follows:  Section \ref{sub:ising} details a mathematical derivation of the Ising model in a grand canonical ensemble for a confined fluid.  Section \ref{sub:thermodyanmic_properties} derives the relevant thermodynamic properties of the adsorbed fluids. Section \ref{sub:bench} explains the benchmarking of the proposed model with GCMC simulations.  Section \ref{sub:phase_transition} discusses the phase transition as a function of pore size based on a double-well potential.  Sec. \ref{sub:phase_diagram} introduces the phase diagram for the adsorbed fluid.  Section \ref{sub:outlook} then discusses how this model may be used in different contexts (e.g., where the confinement effect is significant). Finally,  Sec. \ref{sec:conclusions} presents the conclusions of this paper.

\section{Model}
\label{sec:methods}

\subsection{General assumptions}
We focus on the classical regime, where quantum effects may be disregarded. Consequently, the van der Waals potentials generated by different sources are treated as additive, following the established principles of intermolecular forces \cite{israelachvili2011intermolecular}.
Furthermore, the analysis assumes equilibrium conditions  throughout. This assumption is based on the premise that external conditions, such as temperature and pressure, are time-independent. Given such conditions, macroscopic quantities can be expressed in terms of microscopic average values, distribution functions, or probabilities.

Furthermore,  mean-field theory is used to solve the many-body problem of adsorbents in a potential well of nonuniform depth. This assumption is valid under conditions of low adsorbent concentration because  interactions between adsorbed molecules are negligible at such concentrations. Moreover, the mean-field theory remains valid at and above the saturation point for capillary condensation because the distribution of fluid molecules within  pores becomes uniform, leading to a mean-field effect. However, the mean-field theory may not be accurate in a high-density gas-like regime. In such cases where the  adsorbents are relatively concentrated, a density distribution  around the heterogeneity  creates an anisotropy. However, this effect is not  considered in the current model.

Given these  assumptions, the current work provides a framework for analyzing and understanding the behavior of fluids in confined spaces, particularly in the context of adsorption in MOFs. These assumptions allow for simplified models and calculations, enabling insights into the thermodynamic properties and phase transitions of the adsorbed fluids. 

To clarify the formulation of the model, we briefly revisit the fundamental concepts of Hill's nanothermodynamics \cite{hill2001different} in the context of the current problem (a detailed derivation and discussion are available in Ref. \cite{hill1994thermodynamics}). In our case, fluid confined in the pore of a modeled MOF is in equilibrium with the surrounding bulk fluid. The argon molecules confined within the MOF are the ``system" in this investigation. To understand the thermodynamic characteristics of this system, we subdivide it into an ensemble of \(\eta\) small, equivalent, distinguishable, independent  systems, as shown in Fig.~\ref{fig:figure1}(a). Therefore, assuming the total volume is constant,  the system described here at equilibrium gives 
\begin{gather}
    dE_t = TdS_t + \mu dN_t + \xi d\eta,
    \label{eq:equilibrium_eq1}
\end{gather}
where $E_t$ is the total energy of the system, $T$ is the temperature, $S_t$ is the total entropy of the system, $\mu$ is the chemical potential, $N_t$ are the total number of molecules in the system, \(\xi\) is the subdivision potential, and \(\eta\) is the number of subdivisions. 

Equation (\ref{eq:equilibrium_eq1}) resembles Gibb's equilibrium equation for a two-component system with \(N_t\)  being the number of \textit{molecules}. Since \(\eta\) is the number of subdivisions then the total volume \(V_t = \eta V\), where   \(V\) is the volume of each subdivision. Therefore, we consider the work associated with  varying \(\eta\) at pressure $p$ by adding the work of expansion, \(-p\eta dV\), in Eq.~ (\ref{eq:equilibrium_eq1}) to obtain
\begin{align}
    dE_t = TdS_t -p \eta dV + \mu dN_t + \xi d\eta,
    \label{eq:equilibrium_eq2}
\end{align}
where \(-p\eta \equiv \partial E_t/\partial V\).
We now use Hill's definition of subdivision potential, \(\xi \equiv -\hat{p}V\), where \(\hat{p}\) is the integral pressure \cite{hill1994thermodynamics}. Integrating Eq.~(\ref{eq:equilibrium_eq2}) gives the equilibrium equation for the total system: 
\begin{align}
    E_t = TS_t +\mu N_t - \hat{p}V\eta.      
    \label{eq:equilibrium_eq3}
\end{align}
From here, it is straightforward to show for a grand canonical ensemble that 
\begin{align}
    \hat{p}V = k_\text{B}T \ln  \Xi,
    \label{eq:equilibrium_eq4}
\end{align}
where $k_\text{B}$ is the Boltzmann constant and \(\Xi\) is the grand partition function for the small chosen  system of volume \(V\). 

Hereinafter, we use Hill's notation \cite{hill1994thermodynamics} where the hat (\(\hat{~}\)) denotes the integral intensive thermodynamic function and for any extensive thermodynamic function \(\alpha\), and 
\begin{align}
    \bar \alpha \equiv \frac{1}{V}\int_V \alpha dV
\end{align}
denotes the integral extensive thermodynamic function. In contrast, symbols without hat or  bar represent  differential thermodynamic functions.

\subsection{Ising framework}
\label{sub:ising}
To understand the phase transition, the Ising model plays a crucial role. The inherent complexity in the 3D Ising model coupled with the external non-uniform field presents significant challenges. However, we have made certain assumptions, mentioned in the prior section to obtain an approximate solution. The formulation for the confined fluids is as follows:

Let \(\vec{p} = (p_1,p_2,\dots,p_N)\), \(\vec{q} = (q_1,q_2,\dots,q_N)\) be the momentum and position coordinates, respectively, in phase space for a system of \(N\) molecules confined in a framework creating a potential \(U_{ma}(\vec{q})\). The Hamiltonian \(\mathcal{H}\) is then  
\begin{subequations}
\begin{align}
    \mathcal{H}(\vec{p},\vec{q}) &= \mathcal{K}(\vec{p},\vec{q}) + \mathcal{U}(\vec{p},\vec{q}) 
    \label{eq:chap4_hamilton1}\\
    &= \sum_i^N \frac{p^2_i}{2m} + \sum_i^{N-1}\sum_{j>i}^{N} \Phi(\vec{r}_i-\vec{r}_j) + \sum_i^N U_{ma}(q_i),
    \label{eq:chap4_hamilton2}
\end{align}
\end{subequations} 
where \(\mathcal{K}\) and \(\mathcal{U}\) are kinetic and potential energy contributions to the Hamiltonian, respectively. $m$ is mass of the particle, \(\Phi(\vec r_i - \vec r_j)\) is the potential between molecules \(i\) and \(j\) as a function of the spatial coordinate \(\vec r\). This intermolecular potential can be approximated as a field   defined in terms of the phase-space coordinate \(U_{aa}(\vec{q})\): 
\begin{align}
    \sum_i^NU_{aa}(q_i) \equiv \sum_i^{N-1}\sum_{j>i}^{N} \Phi(\vec{r}_i-\vec{r}_j).
\end{align}

\subsubsection{Canonical ensemble}
 Equation (\ref{eq:chap4_hamilton2}) shows that the kinetic-energy term is independent of the position coordinate \(\vec{q}\) and the potential-energy term is independent of the momentum coordinate \(\vec{p}\). Assuming that the confining framework is stationary and only fluid particles contribute to the kinetic energy, we separate the variables and define the canonical partition function \(\mathbb{Z}\) as the product of the kinetic contribution \(Z_{k-aa}\) and the configurational contribution \(Z_{q}\) [refer Eq.~\ref{eq:can_part_func}]. Additionally, the kinetic energy in turn depends only on the temperature of the reservoir. Therefore, we focus on the solution of the configurational partition.
\begin{align}
    \mathbb{Z} = Z_{k-aa}  Z_{q},
    \label{eq:can_part_func}
\end{align}
where the configurational contribution is 
\begin{align}
    Z_{q} = \frac{1}{V^N}\int_V e^{-\mathcal{U}(q_1, q_2,\dots,q_N)/\tau}dq_1\dots dq_N,
    \label{eq:chap4_conf_part_fun}
\end{align}
with \(\tau = k_\text{B}T\). 
The potential energy is a combination of potentials created by adsorbate-adsorbate interaction and the MOF-adsorbate interaction as described in Eq.~\ref{eq:tot_pe}
\begin{align}
    \mathcal{U}(\vec{q}) = U_{aa}(\vec{q}) + U_{ma}(\vec{q}).
    \label{eq:tot_pe}
\end{align}
\(U_{ma}\) is the potential energy due to the framework at any given position, which implies  
\begin{align}
    U_{ma}(\vec{q}) &= U_{ma}(q_1) + U_{ma}(q_2) + \dots + U_{ma}(q_N).
\end{align}
The configurational partition function [Eq.~(\ref {eq:chap4_conf_part_fun})] is
\begin{align}
    Z_{q} = \frac{1}{V^N}\int_{V} &\exp\{-[U_{ma}(q_1)+\cdots+U_{ma}(q_N) \nonumber\\ 
    +&U_{aa}(q_1)+\cdots+U_{aa}(q_N)]/\tau\}dq_1 \cdots dq_N.
    \label{eq:conf_part_func2}
\end{align}
For the extreme case in the absence of any external field, where \(U_{ma}(q_1,\dots,q_N) = 0\), Eq.~(\ref{eq:conf_part_func2}) resembles the configurational partition function for a bulk fluid. 

The complexity due to the inclusion of an external field may be treated in several ways. For example, Travalloni \textit{et al.} \cite{travalloni2010critical} used the extension of the generalized van der Waals theory to model the confined fluids. Simon \textit{et al.} \cite{simon2014optimizing, simon2017statistical} took as a lattice model the adsorbed gas and the different orientations of the flexible framework and defined a transfer matrix for a rather complex problem. Poluektov \cite{poluektov2015thermodynamic} analyzed the self-consistent field model for classical systems using a one-dimensional perturbation theory. Singh \textit{et al.} \cite{singh2022computing} proposed  decoupling the two types of interactions and approximating the solution. Recently, Dong \textit{et al.} \cite{dong2023thermodynamics} used Gibbs-surface thermodynamics to define the problem in appropriate independent variables and obtained an analytical solution for the special case of confinement between parallel sheets. 

We follow the approach of Singh \textit{et al.} \cite{singh2022computing} of decoupling the interactions and approximating the effect of the external potential using Mayer's \(f\)-functions \cite{mayer1937statistical, mayer1941molecular}. Let \(f_i\) be defined as follows: 
\begin{align}
    f_i \equiv e^{-U_{ma}(q_i)/\tau} - 1.
\end{align}
 Equation (\ref{eq:conf_part_func2}) can then be written as 
\begin{align}
    Z_{q} = \frac{1}{V^N}\int_V e^{-U_{aa}(\vec{q})/\tau}\prod_{i=1}^N(1+f_i)dq_1 \cdots dq_N,
\end{align}
where 
\begin{align}
    \prod_{i=1}^N(1+f_i) = 1 + \sum_{i=1}^N f_i + \sum_{i=1}^{N-1}\sum_{j>i}^Nf_if_j + {O}(f_i^3).
    \label{eq:expansion}
\end{align}
Including the individual molecular terms and ignoring the higher-order terms in Eq.~(\ref{eq:expansion}) yields,
\begin{align}
    Z_{q} =&\, \frac{1}{V^N}\int_V e^{-U_{aa}(\vec{q})/\tau} \left( 1 + \sum_{i=1}^N f_i\right) dq_1 \cdots dq_N
    \label{eq:conf_part_func1}
\end{align}
This expansion is analogous to the first-order fluid-fluid interactions described by  Mayer  \cite{mayer1941molecular}. In this formulation,  Mayer's \(f\)-functions are applied to simplify the heterogeneous interactions and not the fluid-fluid interactions. 

Assuming that, despite the density distribution resulting from the external potential, fluid particles collectively establish a \textit{mean field} to render the analytical solution tractable, we can write 
\begin{align}
    U_{aa}(\vec{q}) = Nu_{aa},
\end{align}
where \(u_{aa}\) is the mean field independent of the positional coordinate in  phase space. Thus, \(u_{aa}\) can be moved outside the integral and Eq.~(\ref{eq:conf_part_func1}) takes the form 

\begin{align}
    Z_{q}=&\,\frac{1}{V^N}\int_V e^{-Nu_{aa}(\vec{q})/\tau}dq_1 \cdots dq_N \nonumber \\
    &+ \frac{e^{-Nu_{aa}(\vec{q})/\tau}}{V} \left(\int_V f_1 dq_1 + \cdots + \int_V f_N dq_N \right).
    \label{eq:conf_part_func}
\end{align}

The first term in  Eq.~(\ref{eq:conf_part_func}) describes the configurational partition function of the bulk phase in the absence of any external field; let it be \(Z_{q-aa}\). In addition, we can define
\begin{align}
    \phi &\equiv \frac{1}{V} \int_V f_i dq_i \nonumber \\
    &= \frac{1}{V} \int_V \left(e^{-U_{ma}(q_i)/\tau} -1 \right) dq_i. 
\end{align}
Using this definition, 
\begin{align}
    Z_{q} \approx Z_{q-aa}\left(1 + N\phi \right),
\end{align}
where \(Z_{q-aa}\) is the configurational partition function of the mean-field bulk fluid.
Putting this back into Eq.~(\ref{eq:can_part_func}) gives
\begin{align}
    \mathbb{Z} &= Z_{k-aa}  Z_{q-aa}\left(1 + N\phi \right) \nonumber \\
    &= \mathbb{Z}_{\text{bulk}}\left(1 + N\phi \right).
    \label{eq:can_part_fun2}
\end{align}
Given that \(U_{ma}\) is a spatially varying potential inside the mesopores, we consider an infinitesimal volume \(dV=dxdydz\) at  coordinate \(q_i = (x,y,z)\). We assume a uniform potential  \(U_{ma}(x,y,z)\) in the infinitesimal volume, which implies that the canonical partition function  for an infinitesimal volume is  
\begin{align}
    \mathbb{Z} = \mathbb{Z}_{\text{bulk}}[1+N\phi(x,y,z)].
    \label{eq:can_part_fun3}
\end{align}

\subsubsection{Grand canonical ensemble}
\label{sub:gcen}

The equilibrium assumption implies that the chemical potential \(\mu^{\text{bulk}}\) of the bulk phase   equals the chemical potential \(\mu^{\text{total}}\) of the argon inside the nanospace. The chemical potential \(\mu^{\text{total}}\) of argon inside the nanospace consists of contributions from other argon inside nanospace (\(\mu^{\text{ads}}\)) and from the framework atoms (\(\mu^{\text{frame}}\)) that form the heterogeneity. Both contributions are made through intermolecular forces, not intramolecular forces:
\begin{subequations}
\label{eq:chem_eq}
\begin{align}
    \mu^{\text{bulk}}(p_{\text{ext}},T) &= \mu^{\text{total}}\\
    &= \mu^{\text{ads}} + \mu^{\text{frame}}  .  
\end{align}
\end{subequations}
Similar to the excess chemical potential defined by Widom \cite{widom1963some}, the chemical potential of the adsorbed phase  combines the intramolecular chemical potential \(\mu^{\text{ads}}\) from  the adsorbed phase and the excess chemical potential \(\mu^{\text{frame}}\) due to the framework. We use the mean value of the interaction energies in the unit cell to calculate the excess chemical potential \(\mu^{\text{frame}}\):
\begin{equation}
    \mu^{\text{frame}} = -\tau\ln\left  \langle \exp \left( \frac{-U_{ma}(\vec{q})}{\tau}\right)\right\rangle .
    \label{eq:widom_insertion}
\end{equation} 
In addition, the unit-cell volume \(V\) is fixed and the temperature \(T\) is controlled externally. Therefore, the modeling is done in a grand canonical ensemble (\(\mu^{\text{ads}}, V, T\)).
Moreover, if a partition function of a system of particles can be obtained, the relevant quantities of interest, such as density, pressure, entropy, and free energy can be derived. To this end, we start the model by defining the grand partition function \(\Xi_{\text{ads}}\) for the adsorbed fluid \cite{hill1994thermodynamics}:
\begin{subequations}
    \begin{align}
    \Xi_{\text{ads}} &= \sum_{N=0}^{\infty}\int\frac{d^{N}\vec{q} d^{N}\vec{p}}{h^{3N} N!} e^{-[\mathcal{H}(\vec{p},\vec{q})-\mu^{\text{ads}} N]/\tau} 
    \label{eq:partition_ads1}\\
    &= \sum_{N=0}^{\infty} \mathbb{Z} e^{N\mu^{\text{ads}}/\tau},
    \label{eq:partition_ads2}
\end{align}
\end{subequations}
where \(h\) is Planck's constant and \(\mathbb{Z}\) is the canonical partition function for an ensemble of \(N\) molecules.
Combining Eqs. (\ref{eq:can_part_fun3}) and  (\ref{eq:partition_ads2}) gives 
\begin{align}
    \Xi_{\text{ads}}(\mu^{\text{ads}},V,T) = \Xi _{\text{bulk}}(\mu^{\text{ads}},V,T)\left(1+\phi\langle N\rangle^{\mu^{\text{ads}}}_{\text{bulk}}\right).
    \label{eq:decoupled}
\end{align}
Under extreme conditions where there is no external potential (i.e., no confinement), the partition function \(\Xi_{\text{ads}}\) simplifies to the partition function \(\Xi_{\text{bulk}}\)  of the bulk fluid. This ensures the consistency and coherence of the equation, particularly in scenarios where confinement effects are negligible.

\subsection{Thermodynamic properties}
\label{sub:thermodyanmic_properties}
When examining intensive thermodynamic functions for a bulk fluid, such as pressure or chemical potential, it is common to assume that the fluid is homogeneous, meaning that the  properties of the fluid are uniform throughout the entire volume under consideration. However, in the context of confined fluids, the presence of heterogeneous interactions introduces nonuniformities in the intensive thermodynamic functions. To address this distribution of properties, Hill introduced both differential and integral thermodynamic functions \cite{hill1961statistical, hill1994thermodynamics}. Differential thermodynamic functions are defined at a specific point in space, indicating their local nature. Conversely, integral thermodynamic functions extend their definition across the entire volume of the system, providing a global characterization.

Given that the phase of any substance is defined for a group of molecules \cite{stanley1971phase, nicolis1989exploring}, understanding the phase of the fluid requires that the integral properties be considered. Therefore, in the exploration of phase transitions, emphasis is placed on global (i.e., integral) thermodynamic properties rather than local (i.e., differential) thermodynamic properties.

Relevant thermodynamic properties such as the grand potential \(\bar \Omega_{\text{ads}}\), the expected number  \(\overline{\langle N_{\text{ads}} \rangle}\) of molecules adsorbed, the pressure  \(\hat{p}_{\text{ads}}\) of the adsorbed phase, entropy \(\bar S_{\text{ads}}\), enthalpy \(\bar H_{\text{ads}}\), Helmholtz free energy \(\bar F_{\text{ads}}\), and Gibbs free energy \(\bar G_{\text{ads}}\) can all be obtained from the grand partition function \(\Xi_{\text{ads}}\), as shown in the following section.
\subsubsection{Grand potential \(\bar \Omega_{\text{ads}}\)}
The grand potential can be obtained from its definition 
\begin{align}
    \Omega_{\text{ads}}(x,y,z) &= -\tau \ln [\Xi_{\text{ads}}(x,y,z)] \nonumber\\
    &=-\tau \ln (\Xi _{\text{bulk}}) - \tau \ln [1+\phi(x,y,z)] \nonumber \\
    &= \Omega _{\text{bulk}} - \tau \ln [1+\phi(x,y,z)].
\end{align}
where $\Omega_{\text{bulk}}$ is the grand potential of the bulk fluid.
Since we are interested in the integral properties inside the MOF pore, we  spatially average the grand potential:
\begin{align}
    \bar \Omega_{\text{ads}} &= \frac{1}{V} \int_V \Omega_{\text{ads}}(x,y,z) dxdydz \nonumber \\
    &= \hat{p}V \equiv \xi. 
\end{align}

\subsubsection{Pressure  \(\hat{p}_{\text{ads}}\) of adsorbed phase}
For any macroscopic system, we can write
\begin{align}
    E_t - TS_t - \mu N_t = -\tau\ln \Xi_t. 
    \label{eq:macro_system}
\end{align}
Combining Eqs. (\ref{eq:equilibrium_eq3}),  (\ref{eq:equilibrium_eq4}), and  (\ref{eq:macro_system}), we obtain the integral pressure \(\hat{p}_{\text{ads}}\) of the adsorbed fluid in terms of the grand partition function: 
\begin{align}
    \xi \equiv \hat{p}_{\text{ads}}V = \tau \ln  \Xi_{\text{ads}} = -\bar \Omega_{\text{ads}}.
\end{align}
Therefore,
\begin{align}
    \hat{p}_{\text{ads}} = -\frac{(\bar \Omega_{\text{ads}})_{\mu,T}}{V}.
    \label{eq:pads}
\end{align}

\subsubsection{Expected number  \(\overline{\langle N_{\text{ads}}\rangle}\) of molecules adsorbed}
The expected number \(\overline{\langle N_{\text{ads}}\rangle}\) of molecules adsorbed can be calculated as follows:
\begin{subequations}
    \begin{align}
        \langle N_{\text{ads}}(x,y,z) \rangle
        &= \frac{\sum_N N\mathbb{Z}_{\text{ads}} e^{\mu^{\text{ads}}N/\tau}}{\Xi_{\text{ads}}} \\
        &=\frac{\langle N\rangle^{\mu^{\text{ads}}}_{\text{bulk}} + \langle N^2\rangle^{\mu^{\text{ads}}}_{\text{bulk}} \phi(x,y,z)}{1 + \langle N\rangle^{\mu^{\text{ads}}}_{\text{bulk}} \phi(x,y,z)},
    \end{align}
    \label{eq:number_ads}
\end{subequations} 
where \(\langle N \rangle^{\mu^{\text{ads}}}_{\text{bulk}}\) is the average number of molecules that would be present in the bulk if the chemical potential were  \(\mu^{\text{ads}}\). Note that the term \(\langle N_{\text{ads}}(x,y,z) \rangle\) is not the actual number of molecules at a given position but the expected number. 
Therefore, the expected total number of molecules in the unit cell is 
\begin{align}
     \overline{\langle N_{\text{ads}} \rangle}  = \frac{1}{V}\int_V \langle N_{\text{ads}}(x,y,z) \rangle dxdydz.
\end{align}
The number of molecules confined within a system is influenced by two factors: intermolecular interactions and heterogeneous interactions. The maximum number of molecules is capped by the volume of the unit cell. These interactions collectively contribute to the effective potential experienced by the adsorbate molecules.  Equation (\ref{eq:decoupled}) uses a decoupling approach to separate these two interactions. The intermolecular interaction among the confined molecules is considered independently, following which the heterogeneous interactions are incorporated as an additional potential term. The volume constraint is accounted for by capping the summation in Eq.~(\ref{eq:number_ads}) at \(N_{\text{max}}\) such that
\begin{align}
    N_{\text{max}} = \left\lfloor \frac{V_{\text{cell}}-V_{\text{m}}-V_{\text{l}}}{b}\right\rfloor ,
\end{align}
where \(\lfloor \cdot \rfloor\) is the floor function, \(V_{\text{cell}}\) is the unit-cell volume, \(V_{\text{m}}\) and \(V_{\text{l}}\) are the volume of the unit cell occupied by the metal and ligand, respectively,  \(b = \sqrt{2}\sigma_a^3\) is the volume of each molecule according to van der Waals theory, and \(\sigma_a\) is the Lennard--Jones  size parameter for adsorbates.

\subsubsection{Other thermodynamic functions}
The following standard thermodynamic relations still apply to the integral values: 
\begin{align}
    \text{Entropy:}~(\bar S_{\text{ads}}) &= \Bigl(\pdv{\bar \Omega_{\text{ads}}}{T}\Bigr)_{p,\mu^{\text{ads}}},\\
    \text{Gibbs free energy:}~(\bar G_{\text{ads}}) &= \mu^{\text{ads}}\overline{\langle N_{\text{ads}} \rangle},\\
    \text{Helmholtz free energy:}~(\bar F_{\text{ads}}) &= \bar \Omega_{\text{ads}} + \mu^{\text{ads}}\overline{\langle N_{\text{ads}}\rangle}, \\
    \text{Enthalpy:}~(\bar H_{\text{ads}}) &= \bar G_{\text{ads}} + T\bar S_{\text{ads}}.
\end{align}

\section{Results and Discussions}
\label{sec:results}
\subsection{Benchmarking}
\label{sub:bench}
\begin{figure}
    \centering    \includegraphics{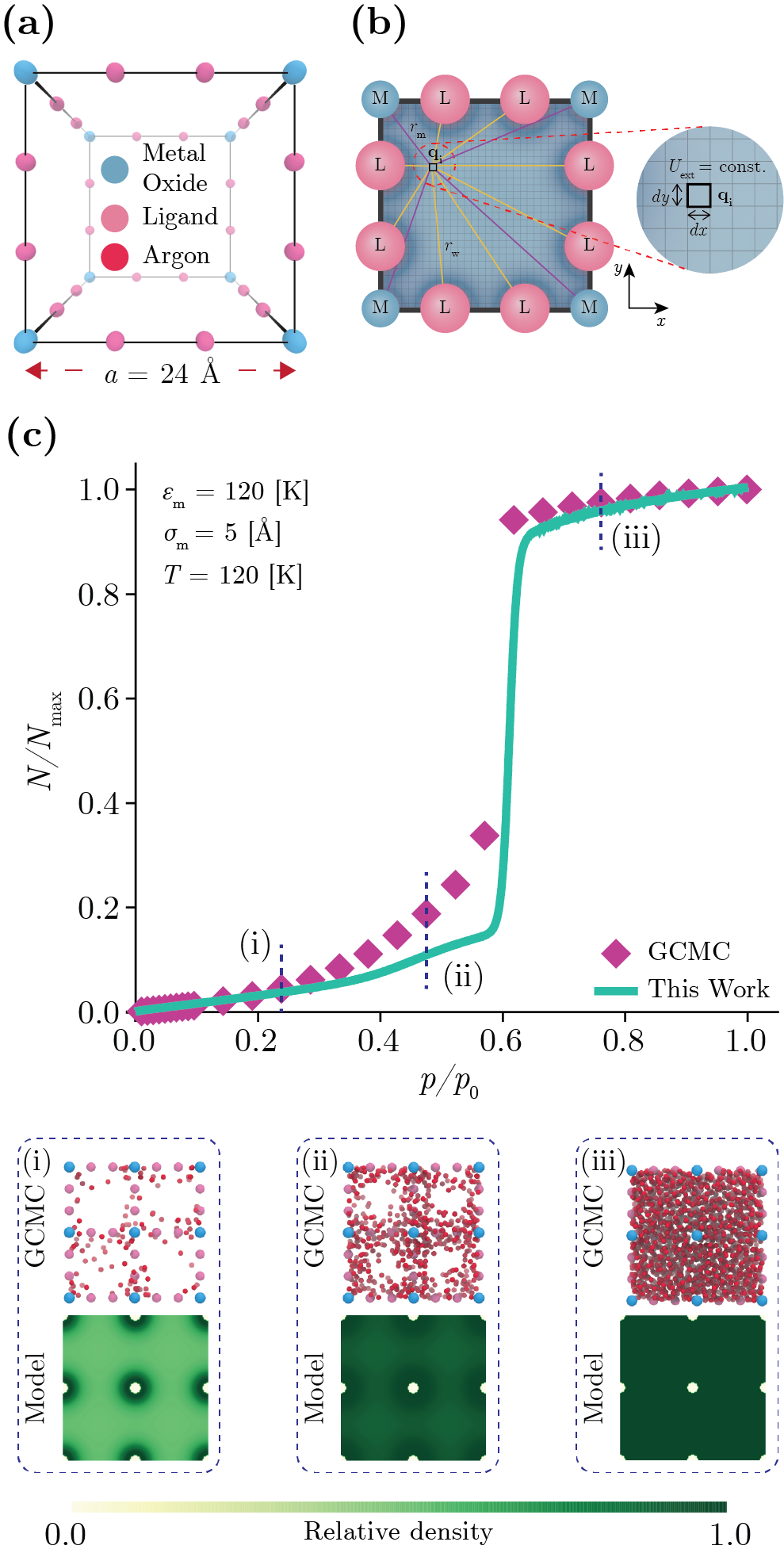}
    \caption{{Benchmarking of the model with GCMC simulations}.    
    (a) Model metal-organic framework (MOF) with metal oxides and ligands; \(a\) is the unit-cell length. (b) Potential-energy distribution inside the unit cell of this MOF. (c) Adsorption isotherm for argon obtained from the model (green line) and the GCMC simulations (pink diamonds). Capillary condensation occurs at relative pressure \(p/p_{\text{0}} \approx 0.6\). The insets (i)--(iii) show the distribution of argon atoms inside the unit cell obtained from the GCMC simulation and the expected relative density distribution (normalized with the maximum local density for better contrast) obtained from the proposed model presented at the corresponding locations in the isotherm. 
    The benchmarking is done for argon adsorption in a model MOF with unit-cell length \(a = 24\) \AA , LJ parameters \(\sigma_m = 5\) \AA ,  \(\varepsilon_m = 120\)~K, and  temperature \(T = 120\)~K.}
    \label{fig:benchmarking}
\end{figure}   
We assume that bulk argon behaves as a van der Waals fluid, so we can write the canonical partition function of the bulk fluid as
\begin{align}
    \mathbb{Z}_{\text{bulk}} = \frac{1}{N!}\left(\frac{V-Nb}{\lambda^3_T}\right)^N \exp \left(\frac{-aN^2}{V\tau}\right),
\end{align}
where \(a\) and \(b\) are the van-der Waals coefficients for argon and \(\lambda_T\) is the thermal de Broglie wavelength \cite{kittel1980thermal}. Based on this assumption various cases with different pore sizes and temperatures were analyzed using an in-house GPU-accelerated Python code (available upon request). 

This proposed model is benchmarked using  a GCMC simulation for a model MOF with a cubic unit cell. Metal oxides occupy the vertices of the cube and ligands are located on the edges of the cube, as shown in Fig.~\ref{fig:benchmarking}(a). To maintain the consistency with the statistical model, we used averaged LJ parameters for metal-oxide and aromatic rings to model this framework \cite{you2022theoretical}. The GCMC simulations were performed using the RASPA \cite{dubbeldam2016} simulation software package. All simulations included a 50\,000-cycle equilibration period and a 100\,000-cycle production run. In these simulations, the structure of all the frameworks is considered rigid; that is, all species of the framework are held fixed at their crystallographic positions. The argon atoms can move in three different ways in the  GCMC simulation: translation, rotation, and swap. The interaction between MOF and argon and between argon atoms was modeled using the LJ potential function and the Lorentz--Berthelot mixing rule. Sample crystallographic information  (.cif) files and the LJ parameters are listed in the electronic supporting information (ESI) \cite{esi} Sec. S1\,A. 

Figure \ref{fig:benchmarking} shows the benchmarking results for the 24 \AA{} unit cell. Figures \ref{fig:benchmarking}(a) and  \ref{fig:benchmarking}(b) show the unit-cell structure and the potential-energy distribution, respectively.   Figure \ref{fig:benchmarking}(c) compares the adsorption isotherm derived from the GCMC simulation  with that produced by the proposed model, showing that the two curves are consistent. Additionally, Fig.~\ref{fig:benchmarking}(c)\ shows in the three lower panels the molecular distribution   obtained through the GCMC simulation along with the density distribution  within the pore at positions (i), (ii), and (iii). The proposed model correctly captures the previously observed trend of layered adsorption \cite{de1981polymer, fukui1999imaging, fukuma2010atomic, page20143, wang2018layered}. Specifically, at lower relative pressures, adsorption predominantly occurs near the heterogeneity, as shown in Fig.~\ref{fig:benchmarking}(c)(i). As the relative pressure increases but before capillary condensation, a distinct layering of adsorbed molecules is evident in Fig.~\ref{fig:benchmarking}(c)(ii). Finally, beyond capillary condensation, the pore becomes saturated with a density distribution  resembling that of the bulk liquid [Fig.~\ref{fig:benchmarking}(c)(iii)]. Additional benchmarking results are presented in Sec. S1. Moreover, Sec. S2 of the ESI \cite{esi} provides a brief parametric investigation that supports the hypothesis made in our previous paper \cite{auti2023effect} that the thermodynamic properties of confined fluids are a function of the confinement parameter \(\Psi \equiv \sigma_{ma}/a\).

 Figure S1 of the ESI \cite{esi} shows that the isotherms for ultrasmall pore size (10 \AA{}) and larger pore sizes (24 \AA{}) are consistent with the results obtained from GCMC simulations. This result is attributed to the assumption of a uniform field generated by the  fluid molecules adsorbed in the cavity. In ultrasmall pores, the variability in the field resulting from argon adsorption at different positions is effectively equivalent to a uniform distribution. Similarly, for larger pore sizes, the distribution of molecules near  adsorption sites remains independent at lower concentrations, thereby creating a uniform field inside the cavity, as depicted in  Fig.~\ref{fig:benchmarking}(c)(i). At higher concentrations (that is, post capillary condensation), argon molecules densely occupy the cavity, resulting in a uniform field, as shown in  Fig.~\ref{fig:benchmarking}(c)(iii). For concentrations between these extremes, the adsorption isotherm slightly deviates from the GCMC values, as shown in   Fig.~\ref{fig:benchmarking}(c)(ii).

Likewise, for the medium pore sizes shown in Figs. S1(b)--S1(e) of the ESI \cite{esi},  the isotherm produced by the proposed model deviates from the GCMC isotherm. This discrepancy arises when the assumption of a uniform field in the cavity is no longer valid. This problem could be addressed through an iterative process, where the density distribution obtained from the proposed model serves as the initial guess. However, this problem is beyond the scope of present paper.

\subsection{Phase transition in confinement}
\label{sub:phase_transition}
\subsubsection{Types of phase transitions}
Since the inception of fullerenes and  3D carbon nanotubes  with cylindrical pores,  the phase-transitions inside these structures have been discussed \cite{ball1993new}. Multiple studies show that freezing of  water confined in these structures may occur continuously or discontinuously \cite{koga2001formation, han2010phase}. The results in this section show that this transition depends on the pore size. Smaller pores produce continuous phase transitions, whereas larger pores produce discontinuous (first-order) phase transitions.

The Helmholtz free energy can be expressed in terms of the canonical partition function [Eq.~(\ref{eq:can_part_fun3})]:
\begin{align}
    F_{\text{ads}}(N,V,T) &=-\tau \ln \bigl[\mathbb{Z}_{\text{bulk}}(1+N\phi)\bigr] \nonumber\\
    &=-N\tau\left\{ \ln  \left[ \frac{(V-Nb)}{\lambda^3_{T}N} \right]+1 \right\} - \frac{N^2a}{V} \nonumber \\
    & - \tau \ln \bigl[1+N\phi(x,y,z)\bigr].
\end{align}
Therefore, we  define the differential chemical potential \(\mu^{\text{ads}}\) and integral chemical potential \(\hat{\mu}^{\text{ads}}\) as follows:
\begin{align}
    &\mu^{\text{ads}} = \left(\pdv{F_{\text{ads}}}{N}\right)_{V,T}, \nonumber \\ &\hat{\mu}^{\text{ads}} = \left(\pdv{\bar {F}_{\text{ads}}}{N}\right)_{V,T} = \left(\frac{\bar {F}_{\text{ads}}}{N}\right)_{V,T}.
    \label{eq:int_diff}
\end{align}

\begin{figure}[h]
    \centering
    \includegraphics{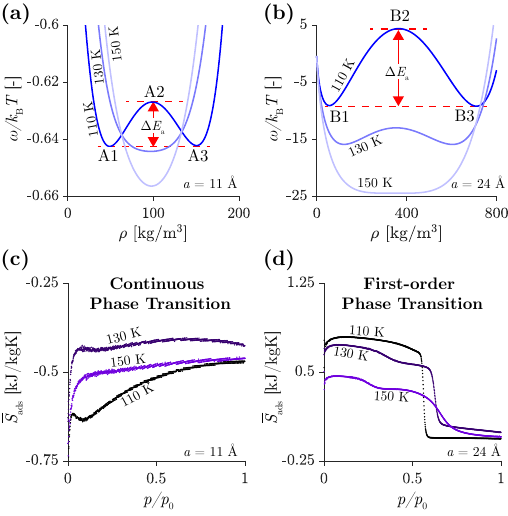}
    \caption{{Types of phase transitions.}
    Potential wells for fluid confined in (a) 11 \AA{} pores and (b) 24 \AA{} pores. Entropy variation is plotted as a function of relative bulk pressure confined in (c) 11 \AA{} pores and (d) 24 \AA{} pores at temperatures ranging from 110 K - 150 K.}
    \label{fig:phase-transition}
\end{figure}

Using the formulation in the canonical ensemble, we  calculate the grand potential \(\omega\) as follows: 
\begin{align}
    \omega = \bar F_{\text{ads}} - N\hat{\mu}^{\text{ads}}_{\text{sat}}.
\end{align}
where \(\hat{\mu}^{\text{ads}}_{\text{sat}}\) is the chemical potential at which capillary condensation occurs for a given temperature.

Figures \ref{fig:phase-transition}(a) and \ref{fig:phase-transition}(b) show the  grand potential thus obtained  plotted as a function of density. For the small pore size (11~\AA), two phenomena occur. First, the double well vanishes at a much lower temperature than for the large pore size, indicating a lower critical temperature for fluids confined in small pores. Consequently, the entropy variation in Fig.~\ref{fig:phase-transition}(c) shows that a continuous phase transition occurs beyond 130 K. Second, the energy barrier to cross the well is less than the thermal noise (\(\Delta E_a \approx 0.015{k}_{\text{B}}T\) for 110 K), implying that the system  spontaneously jumps between the wells. As a result, a minute step occurs in the entropy variation at 110 K at a lower relative pressure, as shown in Fig.~\ref{fig:phase-transition}(c). This  implies that two different phases exist. However, 
\begin{align}
    P_{\text{A}1}/P_{\text{A}2} = \exp (\Delta E_a/k_{\text{B}}T) \approx 1,
\end{align}
implying that the probability of the system being in state A1 or A3 (\(P_{\text{A}1}\)) approximately equals the probability of the system being in state A2 (\(P_{\text{A}2}\)). Therefore, these two phases are practically  indistinguishable. In addition, we  hypothesize that the absence of hysteresis during the adsorption-desorption loop is because the required activation energy is negligible. This hypothesis is consistent with published data that show that the Type-I adsorption isotherm for H\(_2\) adsorption in IRMOF-1 (generally observed for small pores) has no hysteresis \cite{rowsell2004hydrogen}.

In contrast, for the large pore size (24~\AA), the barrier height is significantly larger (\(\Delta E_a \approx 15k_{\text{B}}T\)), highlighting a clear distinction between the ``gas-like adsorbed phase" (i.e., state B1) and the ``capillary condensed phase" (i.e., state B3), as shown in Figs. \ref{fig:phase-transition}(b) and  \ref{fig:phase-transition}(d). However, at 150~K, the double well completely vanishes, and the adsorbed fluid exhibits a single well, which is characteristic of the supercritical bulk fluid, suggesting that the confined fluid has a critical point. Furthermore, based on the  activation energy required for  capillary condensation, we  hypothesize that  hysteresis occurs during the adsorption-desorption process for large pores. As  seen for the adsorption of H\(_2\) in MOF-253 with a Type-V adsorption isotherm (generally observed for large pores), significant hysteresis occurs in the adsorption-desorption loop \cite{yaghi2012hydrogen}.

\subsubsection{Capillary condensation}
\begin{figure}[h]
    \centering
    \includegraphics{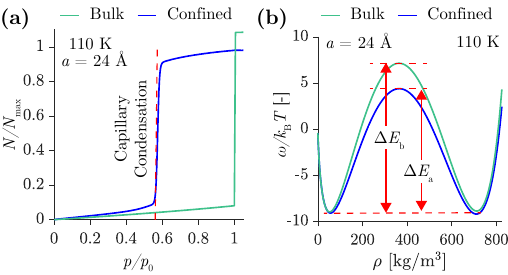}
    \caption{{Energy barrier for capillary condensation.}
    (a) Isotherm of relative number density vs relative pressure for adsorbed argon and the bulk argon. Adsorbed argon condenses at a lower pressure than bulk argon (\(p/p_{\text{0}}\approx0.6\)). $N_{\text{max}}$ is the maximum number of molecules that can be adsorbed. (b) The energy barrier for the phase transition of bulk argon is higher than that for confined argon (\(\Delta E_b > \Delta E_a\)).}
    \label{fig:whycapcond}
\end{figure}

 Figure \ref{fig:whycapcond}(a) shows that the phase transition for the adsorbed fluid occurs at a lower relative pressure than  for the bulk phase transition. To elucidate this, we  compare  the nucleation of droplets in a bulk fluid with that in a confined fluid.

As with macroscale condensation, capillary condensation starts with the heterogeneous nucleation of drop clusters that subsequently grow to form droplets \cite{bocquet1998moisture, restagno2000metastability, maeda2002nanoscale}. Consistent with classical nucleation theory \cite{debenedetti1996metastable}, the  nucleation rate depends on the nucleation barrier \(\Delta G^*\), which is the difference between the interface free energy and the fluid free energy. A lower nucleation barrier corresponds to a higher nucleation rate.
Figure \ref{fig:whycapcond}(b) compares the energy barrier \(\Delta E_b\) for the bulk fluid with that for the adsorbed fluid (\(\Delta E_a\)). The result shows that the free energy for the adsorbed fluid is notably lower than that for the bulk fluid, implying that the free-energy barrier \(\Delta G^*\) for the confined drop nucleation is less than that for nonconfined drop nucleation. Consequently, at a given temperature, the condensation pressure for the adsorbed fluid is less than that for the bulk fluid, as shown in Fig.~\ref{fig:whycapcond}(a).

\subsection{Phase diagram of adsorbed fluid}
\label{sub:phase_diagram}
\begin{figure*} 
    \centering
    \includegraphics[width=0.8\textwidth]{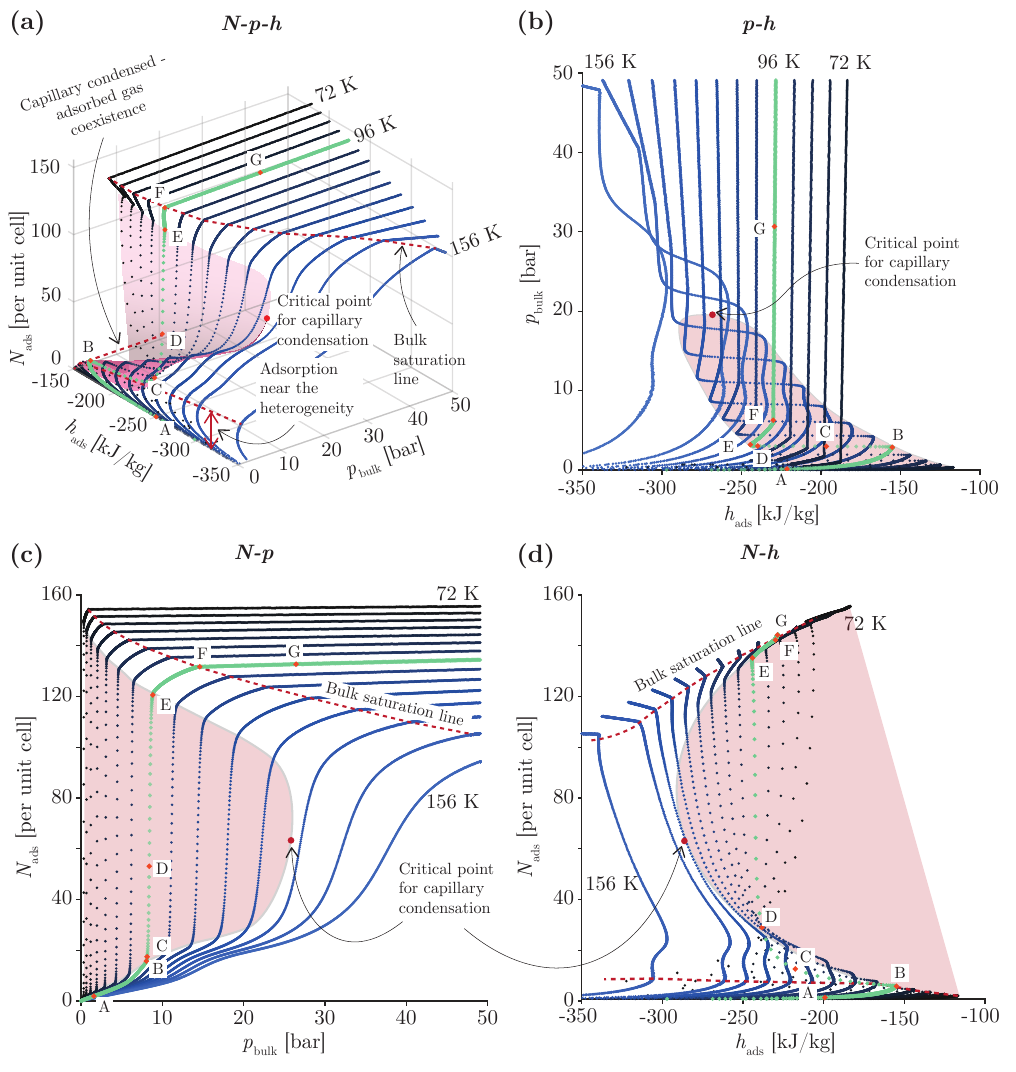}
    \caption{{Phase diagram for argon adsorbed in a metal-organic framework (MOF).} 
    The integral properties are plotted in the form of a phase diagram for the adsorbed fluid. Different perspectives of the (a) 3D \(N\text{-}p\text{-}h\) diagram are plotted in the insets: (b) \(p\text{-}h\), (c) \(N\text{-}p\), and (d) \(N\text{-}h\). The red highlighted areas are the phase-coexistence region constructed using the boundaries of discontinuities on either end of each isotherm. Using this coexistence region, a critical point for capillary condensation is depicted with a red dot. The isotherms are plotted every 6 K and one isotherm in each inset is highlighted with green, to show the shape of the isotherm. Each isotherm follows the direction A\(\rightarrow\)B\(\rightarrow\)C\(\rightarrow\)D\(\rightarrow\)E\(\rightarrow\)F\(\rightarrow\)G. }
    \label{fig:phase-diagram}
\end{figure*}
From an application perspective, the phase diagram is an essential tool for designing engineering processes. For example, Lilley and Prasher \cite{lilley2022ionocaloric} presented a qualitative phase diagram for crystallization of salts in an ionocaloric refrigeration cycle. In a previous paper \cite{shamim2022concept}, we introduced the concept of a 3D phase diagram as a valuable tool for the design of hybrid compression-adsorption heat-pump  systems. The statistical model discussed earlier provides a framework for constructing such a phase diagram. Previous research  has also aimed to construct phase diagrams for confined fluids. For example, numerous attempts have been made to construct such phase diagrams by using Monte Carlo simulations \cite{page1996monte, radhakrishnan2000effect, radhakrishnan2002global, de2007phase}. However, such simulations are computationally demanding, limiting the number of adsorption isotherms that can be generated. Consequently,  acquiring the requisite thermodynamic properties  to create a phase diagram for adsorbed fluids is a significant challenge. 

Radhakrishnan \textit{et al.} \cite{radhakrishnan2002global} solved this problem by applying umbrella sampling and bias potentials to compute the system's free energy. However, such an approach necessitates \textit{a priori} knowledge of the process, and the convergence of their method depends heavily  on the selected collective variables.
Lum and Chandler \cite{lum1998phase} addressed this issue within the framework of statistical mechanics, albeit for a specific scenario, by deriving a phase diagram for vapor confined within a cylindrical pore. Unfortunately, this approach overlooks the nonuniform characteristics inherent in the external field at this length scale, thus lacking generalizability.
In contrast, Travalloni \textit{et al.} \cite{travalloni2010critical} constructed a phase diagram employing square-well potentials to account for external heterogeneity. Nevertheless, this particular study does not consider the excess chemical potential induced by external interactions, leading to unrealistically high pressures during capillary condensation.

Figure \ref{fig:phase-diagram}(a) illustrates the 3D phase diagram of adsorbed argon within a 24~\AA{} model of an MOF. The three axes correspond to distinct state variables: the external pressure is denoted  \(p_{\text{bulk}}\), the enthalpy per unit mass of adsorbed argon is denoted  \(h_{\text{ads}}\), and the number of argon molecules adsorbed per unit cell of the model MOF is denoted  \(N_{\text{ads}}\). Figures \ref{fig:phase-diagram}(b)--\ref{fig:phase-diagram}(d) portray the projections of this phase diagram from three distinct orientations, generating the \(p\text{-}h\), \(N\text{-}p\), and \(N\text{-}h\) diagrams. Notably, the isotherms displayed contain discontinuities up to a certain temperature, indicating that two distinct phases coexist. The connection of these discontinuous points yields saturation lines, which are accentuated in red. The highlighted area signifies the coexistence region corresponding to the \textit{capillary condensed phase} and the \textit{gas-like adsorbed phase}. In this phase diagram, A\(\rightarrow\)B adsorption occurs near the heterogeneity-forming layered structure inside the pore, whereas B\(\rightarrow\)C\(\rightarrow\)D\(\rightarrow\)E depicts the coexistence region. E\(\rightarrow\)F shows that the capillary condensed liquid density increases with pressure. Finally, beyond the bulk saturation line F\(\rightarrow\)G, the pore is completely filled. 

Furthermore, a detailed examination of the \(p\text{-}h\) diagram  in Fig.~\ref{fig:phase-diagram}(b) shows that, at low pressure, the enthalpy of the adsorbed fluid  contrasts with that of the bulk fluid. As the pressure decreases, the magnitude of the enthalpy of the adsorbed fluid increases. This phenomenon can be understood by considering the occurrence of layered adsorption near the metallic heterogeneous site. The heightened cohesive interaction with the adsorption site  liberates additional energy, increasing the  enthalpy at lower pressures. ESI Fig.~S8 shows the bulk argon \(p\text{-}h\) diagram.

An important observation from the phase diagram is the absence of discontinuities in the isotherms beyond a specific temperature, resembling the behavior of bulk fluids. This temperature is denoted  the \textit{critical point for capillary condensation}. Beyond this critical point, capillary condensation for the adsorbed fluid ceases, resulting in a lack of stepwise behavior in the adsorption isotherm. Notably, the critical point for capillary condensation is positioned at a lower temperature compared with the bulk critical point of argon (151~K, 48.5~bar). This difference is attributed to the excess chemical potential of the adsorbed fluid vis-\'a-vis the bulk fluid, a consequence of heterogeneous interactions. Similar results of reduction in critical pressure of the liquid-liquid phase transition have been observed for water in a salt solution, where the salt ions act as the heterogeneity \cite{gallo2016water,perin2023phase}.  

This phase diagram provides a basis to understand the phase transition of confined fluids. Note that a critical order parameter analysis for the ``gas-like'' adsorbed phase to ``capillary-condesed'' liquid phase and the analytical construction of the co-existence region still remains to be addressed. 

\section{Outlook and Conclusions}
\subsection{Outlook}
\label{sub:outlook}
This paper proposes a method to predict the integral pressure inside  MOF pores (Sec. \ref{sec:methods}). Figure S3(d) of the ESI \cite{esi} shows the calculated  integral pressure. \(\hat{p}_{\text{ads}}\) is significantly greater than the bulk pressure \(p\) and  jumps discontinuously upon crossing the pore boundary. Therefore, we  define a disjoining pressure \(\Pi_d\) \cite{derjaguin1936properties, israelachvili2011intermolecular} such that
\begin{equation}
    \Pi_d \equiv \hat{p}_{\text{ads}} - p_{\text{bulk}}.
\end{equation}

In a similar way, when considering a constant-pressure ensemble (\(NPT\)), we  establish the integral chemical potential such that \(\xi \equiv \hat{\mu}^{\text{ads}}N\).
Subsequently, a disjoining chemical potential \(M_d\) \cite{dong2023nanoscale} is defined as
\begin{equation}
    M_d \equiv \hat{\mu}^{\text{ads}} - \mu ^{\text{bulk}}.
\end{equation}

The proposed formulation can then be extended in other contexts where wall effects and nonuniform external fields are significant, enabling us to calculate these disjoining quantities to explain phenomena other than adsorption. 

For example, numerous attempts have been made to explain the stability of surface nanobubbles using the Young--Laplace equation and considering external conditions such as surface charge \cite{shi2021probing, tan2021stability} and contact-line pinning due to hydrophilic and hydrophobic heterogeneities \cite{maheshwari2016stability, tan2021stability}. However, a consensus regarding the exact factors contributing to surface-bubble stability remains elusive. The effect of surface heterogeneity  clearly explains the excessive pressure inside these bubbles in terms of a disjoining pressure, offering a plausible explanation for their stability \cite{svetovoy2016effect}.

Similarly, experimental observations reveal that the nucleation temperature for heterogeneous boiling is consistently lower than predicted by classical nucleation theory \cite{witharana2012bubble, chen2019measuring, paul2020single}. Numerous models \cite{lutsko2018systematically, abyzov2014heterogeneous, magaletti2020unraveling} and complex simulations \cite{shen2001density,kimura2002molecular,lutsko2011density} have been proposed to explain this disparity between experimental  and theoretical results. Many such models consider liquid-vapor surface tension through the Young--Laplace equation and liquid-solid surface tension through contact-angle variation. However,  to the best of our knowledge, none of these models account for the wall effect, leading to a disjoining pressure between vapor clusters and the liquid, thereby lowering the free-energy barrier for phase transition, as illustrated  in Fig.~\ref{fig:whycapcond}(b). The lower free-energy barrier in turn  explain the lower the nucleation temperature  of heterogeneous boiling \cite{majumdar1999instability}. 

Additionally, in the context of water transport through nanoscale osmotic membranes,  a recent study \cite{song2021true} argued that  no chemical potential gradient exists for transport within the membrane. However,  the nanometer-scale  membrane porosity allows surface effects to trigger  a disjoining chemical potential, thereby creating a chemical-potential gradient that drives the transport process.  Considering how surface effects have a heightened impact on the nanoscale will allow a better understanding of water transport in osmotic membranes.
\subsection{Conclusions}
\label{sec:conclusions}
In conclusion, our investigation offers a statistical approach to address the 3D Ising model of phase transitions in confined fluids, producing  reasonably  accurate results. By  applying the proposed model and integrating Hill's theory of nanothermodynamics, we  derive both the differential (local) and integral (global) thermodynamic properties of the adsorbed fluid. The proposed model has practical utility for predicting the behavior of adsorbed fluids within porous structures, facilitating the design of materials tailored to specific requirements. The key insights derived from this model are outlined as follows:

First, the nature of the phase transition in the confined fluid is determined by the extent of confinement, specifically the pore size. In small pores, the activation-energy barrier for phase transition is approximately  0.01~\(k_\text{B}T\), significantly less than the thermal noise. Consequently, the phase transition occurs spontaneously. Additionally, the unstable and metastable states, while theoretically existent, are practically indistinguishable from the stable state. Conversely, in the case of large pores, the activation-energy barrier for phase transition is of the order of  10~\(k_\text{B}T\), clearly distinguishing between the gas-like adsorbed phase and the capillary condensed phase.

Second, owing to additional interactions with the surface, the free-energy barrier for phase transitions in confined fluids is lower than in bulk fluids. This reduced energy barrier implies that condensation inside  MOF pores occurs at a lower pressure for a given temperature, explaining the lower capillary condensation pressure. 

Finally, the model proposed in this paper  consolidates the integral thermodynamic properties in the form of a phase diagram for confined fluids. The phase diagram resembles  the bulk fluid phase diagram except for the higher enthalpy released at lower pressure and the lower critical temperature and pressure due heterogenous interactions. 

\begin{acknowledgments}
We acknowledge funding by JST, CREST (Grant No. JPMJCR17I3).
\end{acknowledgments}

\appendix

\bibliography{bibliography}

\end{document}


\title{Supplementary information for the Statistical modeling of equilibrium phase transition confined fluids}
\author{Gunjan Auti} 
\email{gunjanauti@thml.t.u-tokyo.ac.jp}
\affiliation{
Department of Mechanical Engineering, The University of Tokyo, \\7-3-1 Hongo, Bunkyo-ku, Tokyo 113-8656, Japan}

\author{Soumyadeep Paul}
\affiliation{
Department of Mechanical Engineering, The University of Tokyo, \\7-3-1 Hongo, Bunkyo-ku, Tokyo 113-8656, Japan}
\affiliation{
Department of Mechanical Engineering, Stanford University, Building 530, 440 Escondido Mall, Stanford, California 94305, USA}

\author{Wei-Lun Hsu}
\affiliation{
Department of Mechanical Engineering, The University of Tokyo, \\7-3-1 Hongo, Bunkyo-ku, Tokyo 113-8656, Japan}

\author{Shohei Chiashi}
\affiliation{
Department of Mechanical Engineering, The University of Tokyo, \\7-3-1 Hongo, Bunkyo-ku, Tokyo 113-8656, Japan}

\author{Shigeo Maruyama}
\affiliation{
Department of Mechanical Engineering, The University of Tokyo, \\7-3-1 Hongo, Bunkyo-ku, Tokyo 113-8656, Japan}

\author{Hirofumi Daiguji}
\email{daiguji@thml.t.u-tokyo.ac.jp}
\affiliation{
Department of Mechanical Engineering, The University of Tokyo, \\7-3-1 Hongo, Bunkyo-ku, Tokyo 113-8656, Japan}%

\maketitle
\section{Bench-marking}
\label{sec:si_bench}

\subsection{Crystallographic information}
\label{sub:si_cif}
The GCMC simulations for the bench-marking of the model presented in main paper are performed using RASPA \cite{dubbeldam2016}. The structure of the model MOF used for these simulations is shown in the Fig.~2(a) of the main manuscript. The crystallographic information file (.cif) is presented in the Listing 1. To maintain the consistency between the statistical model and the GCMC simulations, coarse-grained model for the ligands have been used, Lennard-Jones (L-J) parameters for each ligand as: 
$\sigma_\text{l} = 5.5$ [\AA] and $\varepsilon_\text{l} = 600$ [K] as reported in the literature \cite{you2022theoretical}. 
\begin{lstlisting}[caption={.cif file for the model MOF ($a$ = 24 \AA)}]
data_model_mof
_cell_length_a   24.0
_cell_length_b   24.0
_cell_length_c   24.0
_cell_angle_alpha   90.0
_cell_angle_beta    90.0
_cell_angle_gamma   90.0
_symmetry_cell_setting          triclinic
_symmetry_space_group_name_Hall 'P 1'
_symmetry_space_group_name_H-M  'P 1'
_symmetry_Int_Tables_number     1
_symmetry_equiv_pos_as_xyz 'x,y,z'
loop_
_atom_site_type_symbol
_atom_site_label
_atom_site_fract_x
_atom_site_fract_y
_atom_site_fract_z
M       M1   	0.0000  0.0000  0.0000
L   	L1_1 	0.0000  0.0000  0.3334
L   	L1_2 	0.0000  0.0000  0.6667
L   	L2_1   	0.0000  0.3334  0.0000
L   	L2_2   	0.0000  0.6667  0.0000
L   	L3_1  	0.3334  0.0000  0.0000
L   	L3_2  	0.6667  0.0000  0.0000
\end{lstlisting}

\newpage
\subsection{Additional bench-marking results}
\label{sub:si_add_bench}
\begin{figure*}[h]
    \centering
    \includegraphics{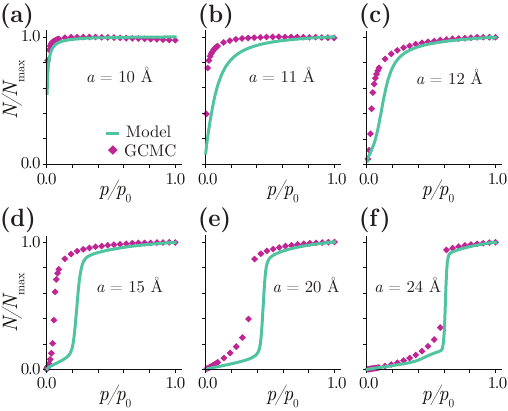}
    \caption{\textbf{Additional bench-marking: Adsorption isotherm}\\
    Additional results for the benchamarking of the adsorption isotherms at 120 K for pore sizes ranging from 10\AA{} to 24\AA. The purple diamonds represent the results from GCMC simulations and the green line represents the isotherms obtained from the model.}
    \label{si_fig:bench_iso}
\end{figure*}

Along with the bench-marking results presented in Fig.~2(c) of the main manuscript, here we present the adsorption isotherms and enthalpy of adsorption results for pore sizes ranging from 10 \AA{} to 24 \AA{}. For ultra-small pore and larger pore shown in Fig.~\ref{si_fig:bench_iso}(a) and (f), respectively, the adsorption isotherm shows good agreement with the GCMC results. However, for the medium sized pores shown in Figs.~\ref{si_fig:bench_iso}(b)-(e), the isotherm obtained from the model slightly deviates from the GCMC results. This disparity can be attributed to the mean field assumption. 

The enthalpy of adsorption derived from the statistical model exhibits an higher values at lower pressures in contrast to the values calculated from the GCMC simulations. This discrepancy arises from the fact that, in the present model, fractional molecules (less than 1) can be adsorbed at lower pressures to ensure mathematical stability. Nevertheless, this issue can be rectified by imposing appropriate constraints on the lower limit of adsorption. Beyond this consideration, overall agreement is observed between the values obtained from the model and the GCMC simulations. The enthalpy of adsorption abruptly increases near saturation, which is consistent with the literature \cite{torres2017behavior}. 

\begin{figure*}
    \centering
    \includegraphics{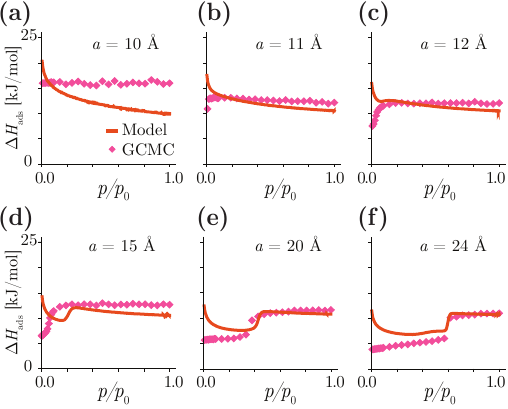}
    \caption{\textbf{Additional bench-marking: Differential enthalpy of adsorption}\\
    Additional results for the bench-marking of the differential enthalpy of adsorption at 120 K for pore sizes ranging from 10~\AA{} to 24~\AA. The pink diamonds represent the results from GCMC simulations and the red line represents the isotherms obtained from the model.}
    \label{si_fig:bench_enth}
\end{figure*}

\newpage
\section{Parametric Study}
\label{sec:si_para}
Based on the model presented in the paper, a short parametric study is presented in this section. The base case for this study is with unit cell size $a = 24$~\AA, L-J parameters $\sigma_\text{m} = 5$~\AA{}, $\varepsilon_\text{m} = 120$~K, and the temperature $T = 120$~K.

\subsection {Temperature response}
\begin{figure*}[h]
    \centering
    \includegraphics[width=\textwidth]{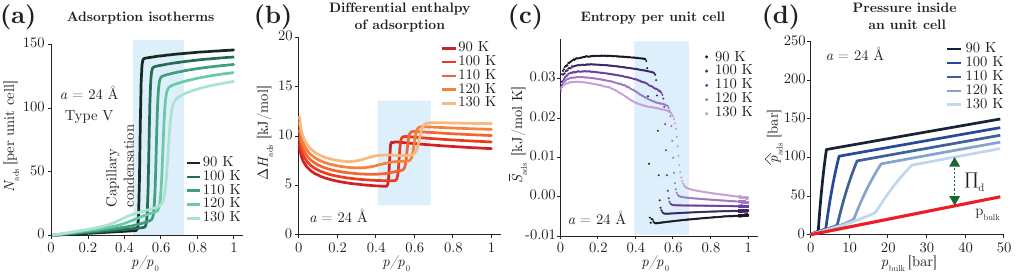}
    \caption{\textbf{Thermodynamic properties - temperature response} (a = 24 \AA)\\
    (a) Adsorption isotherms obtained using the model, the highlighted region shows the relative pressure at which capillary condensation occurs. (b) Differential enthalpy of adsorption $\Delta H_\text{ads}$, at lowest relative pressure $p/p_\text{0} < 0.1$, owing to stronger interaction with the MOF walls, enthalpy released is higher and the abrupt jump in $\Delta H_\text{ads}$ (the blue highlighted region), signifies the latent heat of capillary condensation. (c) The entropy of the adsorbed fluid decreases abruptly after the capillary condensation (the blue highlighted region) showcasing clear distinction between the phases. (d) The integral pressure inside the pore $\hat{p}_\text{ads}$, is significantly higher than bulk pressure. The pressure difference across the pore boundary is known as the disjoining pressure.}
    \label{si_fig:results-24A}
\end{figure*}
The temperature response is studied by considering temperature variation in the base case. The obtained thermodynamic functions are plotted in the Fig.~\ref{si_fig:results-24A}. At lower pressures, below the the capillary condensation pressure, the amount of gas adsorbed at the same relative pressure is higher for higher temperature. In addition, the relative capillary condensation pressure increases with increase in temperature. This is consistent with the studies in literature for adsorption of water in MOFs \cite{furukawa2014water}. The magnitude of enthalpy of the adsorbed fluid increases with the temperature, therefore differential enthalpy of adsorption (difference between specific enthalpy adsorbed fluid and bulk fluid for a given relative pressure) increases with temperature. The temperature response for entropy of adsorbed fluid resembles the temperature response for the bulk fluid, except the relative condensation pressure. Moreover, with respect to the bulk fluid, the entropy of confined fluid is lesser for all relative pressure, owing to more ``organized" arrangement inside the confinement. Finally, the integral thermodynamic pressure can be calculated from the statistical model. The difference between integral pressure inside the pore and the bulk pressure is called as the disjoining pressure ($\Pi_\text{d}$). As the temperature increases the disjoing pressure decreases as the density of the fluid inside the pore decreases.

\subsection{Variation in pore size}
\begin{figure*}
    \centering
    \includegraphics[width=\textwidth]{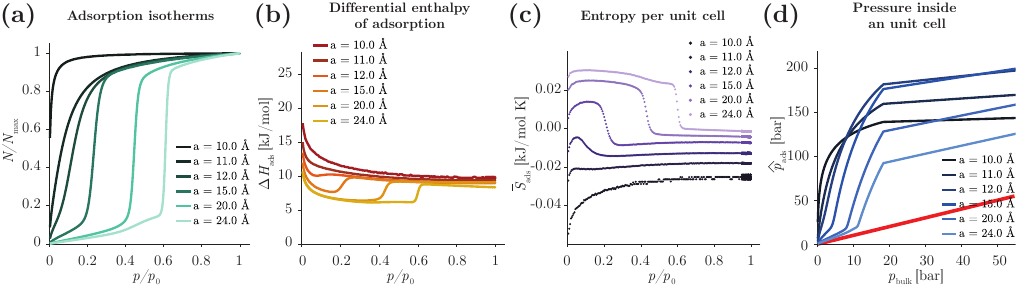}
    \caption{\textbf{Thermodynamic properties - variation of pore size} (T = 120 K)\\
    Variation of thermodynamic properties obtained for pore sizes ranging from 10~\AA{} to 24~\AA{} at 120~K. (a) Type of the adsorption isotherm changes from IUPAC Type I to Type IV to Type V as the pore size increases. (b) Differential enthalpy of adsorption is higher for a smaller pore and progressively decreases as the pore size increases. (c) The entropy variation with respect to relative pressure is shown for different pore sizes. (d) The integral thermodynamic pressure inside different pore sizes with respect to the bulk pressure. }
    \label{si_fig:results-120K}
\end{figure*}

The change in the shape of adsorption isotherm as shown in Fig.~\ref{si_fig:results-120K}(a) indicates the type of phase transition inside the confinement. For smaller pores a continuous increase in density is observed, whereas for larger pores a step-wise increase in the density is observed which is a characteristic of the first-order phase transition. The differential enthalpy of adsorption progressively decreases with the increase in the pore size signifying the strength of interactions with heterogeneity \cite{auti2023effect} as shown in Fig.~\ref{si_fig:results-120K}(b). 

\begin{figure*}
    \centering
    \includegraphics[width=\textwidth]{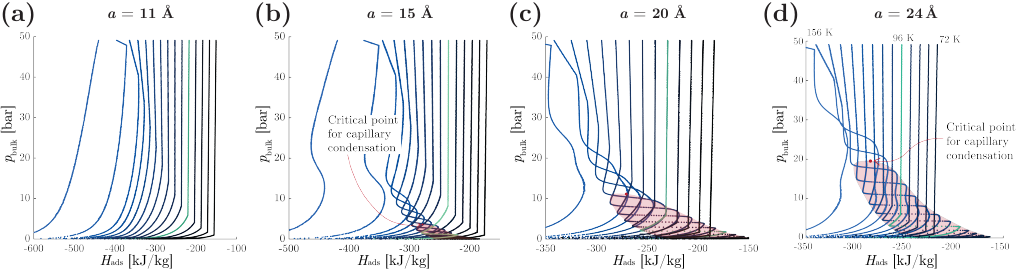}
    \caption{\textbf{p-h Diagram - variation of pore size}\\
    The phase diagram of fluid confined in pore with unit cell size (a) 11~\AA{}, (b)15~\AA{}, (c)20~\AA{}, and (d) 24~\AA{}. The red highlighted region shows the co-existence region for ``gas-like" adsorbed phase and the ``capillary condensed" adsorbed phase. }
    \label{si_fig:si_phase_diagram}
\end{figure*}

The entropy variation (Fig.~\ref{si_fig:results-120K}(c)) shows that the type of phase transition for confined fluid depends on the pore size. For smaller pores a continuous phase transition is observed, whereas for larger pores, a discontinuity in the entropy indicates first-order phase transition. Finally, the variation in integral thermodynamic pressures with respect to pore sizes is shown in Fig.~\ref{si_fig:results-120K}(d). For pore, the integral pressure at lower bulk pressures ($p_\text{bulk} <$ 10~bar ) is higher, similar to the density variation. However, due to the volume constraints, the maximum density in smaller pores is limited, therefore the integral thermodynamic pressure saturates at this value.

Fig.~\ref{si_fig:si_phase_diagram} shows the variation in phase diagram of the fluid confined in different pore sizes. For a smaller pore (11~\AA) as shown in the inset (a), the co-existence region does not exist, which similar to the trans-critical region of the bulk fluid phase diagram. Moreover, it can be observed that the critical temperature and pressure increases as the pore size increases.

\subsection{Variation in the energy parameter of the metal atom}
\begin{figure*}
    \centering
    \includegraphics[width=\textwidth]{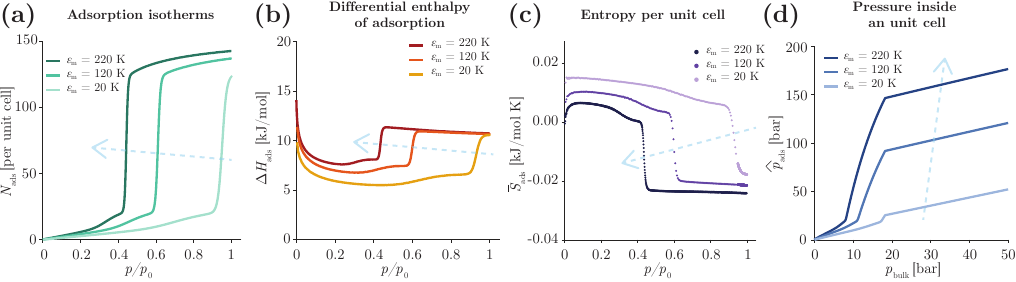}
    \caption{\textbf{Thermodynamic properties - variation of L-J parameter $\varepsilon_\text{m}$} ($\sigma_m = 5$~\AA) \\
    Variation of thermodynamic properties obtained for energy parameter of metal atom ranging from 20~K to 220~K for a given $\sigma_\text{m} = $ 5~\AA{} and temperature 120~K.
    (a) The adsorption isotherm show that for higher energy parameter the relative pressure for capillary condensation is lesser, (b) The enthalpy of adsorption for higher energy parameter is higher, (c) The entropy variation for variation in the energy parameter, and (d) The integral pressure inside the unit cell.}
    \label{si_fig:results-s5a}
\end{figure*}
As the energy parameter for the metal atom increases the magnitude of chemical potential of the framework ($\mu^\text{frame}$) increases. Therefore, the value of chemical potential of the system argon molecules becomes more negative, therefore the capillary condensation occurs at a lower relative pressure as shown in Fig.~\ref{si_fig:results-s5a}(a). For $\varepsilon_\text{m} = 20$~K, the effect of heterogeneity is minimal therefore the relative pressure for capillary condensation is closer to unity (i.e. the bulk value). Similar trend can be observed from the entropy variation depicted in Fig.~\ref{si_fig:results-s5a}(c)

In addition, owing to the higher energy released during the interaction with heterogeneity, the differential enthalpy of adsorption is higher for higher $\varepsilon_\text{m}$, as shown in Fig.~\ref{si_fig:results-s5a}(b). Likewise, the disjoining pressure is higher for higher $\varepsilon_\text{m}$ as shown in Fig.~\ref{si_fig:results-s5a}(d).

\subsection{Variation in the size of the metal atom}
\begin{figure*}
    \centering
    \includegraphics[width=\textwidth]{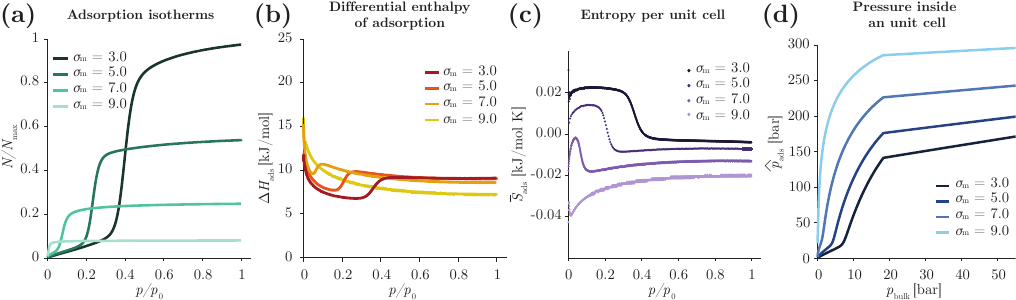}
    \caption{\textbf{Thermodynamic properties - variation of L-J parameter $\sigma_\text{m}$} ($\varepsilon_\text{m} =~$120~K)\\
    Variation of thermodynamic properties obtained for size parameter of metal atom ranging from 3~\AA{} to 9~\AA{} for a given $\varepsilon_\text{m} = $ 120~K and temperature 120~K.
    (a) The shape of adsorption isotherm show that for larger metal atom shows a Type I adsorption isotherm whereas for smaller metal atom Type V adsorption isotherm is observed, (b) The enthalpy of adsorption for all the cases is more or less equal except the relative pressure of phase transition differs, (c) Entropy variation: Largest metal atom, continuous change is entropy, whereas for smaller metal atom a discontinuity in entropy variation, and (d) The integral pressure inside the unit cell is higher for larger value of $\sigma_\text{m}$. }
    \label{si_fig:results-e120k}
\end{figure*}
The adsorption isotherms show a variation in the shape as shown in Fig.~\ref{si_fig:results-e120k}(a) due to available space in the pores of the metal-organic structure, as the value of $\sigma_\text{m}$ increases the volume of the available pore decreases. Enthalpy of adsorption shows little to none variation with respect to the size of the metal atom as shown in Fig.~\ref{si_fig:results-e120k}(b). The entropy variation shown in Fig.~\ref{si_fig:results-e120k}(c), resembles to the entropy variation of different pore sizes depicted in Fig.~\ref{si_fig:results-e120k}(c). The larger metal atom size shows continuous phase transition and it becomes first-order phase transition for smaller metal atoms. 
In addition, the disjoining pressure is higher for larger metal atoms due to higher density of adosrbed fluid, whereas its lesser for smaller metal atoms due to the lower density. 

The variation in the thermodynamic properties of the confined fluids with respect to the size of metal atom (Fig.~\ref{si_fig:results-e120k})  resembles the variation in the thermodynamic properties with respect to the pore size (Fig.~\ref{si_fig:results-120K}). This confirm the hypothesis we presented in an earlier paper \cite{auti2023effect}, that the adsorption isotherms and the properties of the adsorbed fluids are a function of confinement parameter ($\Psi \equiv \sigma_{ma}/a$), where $a$ is the unit cell length.  

\section{Comparison with bulk argon}
\begin{figure}[h]
    \centering
    \includegraphics[width=0.8\textwidth]{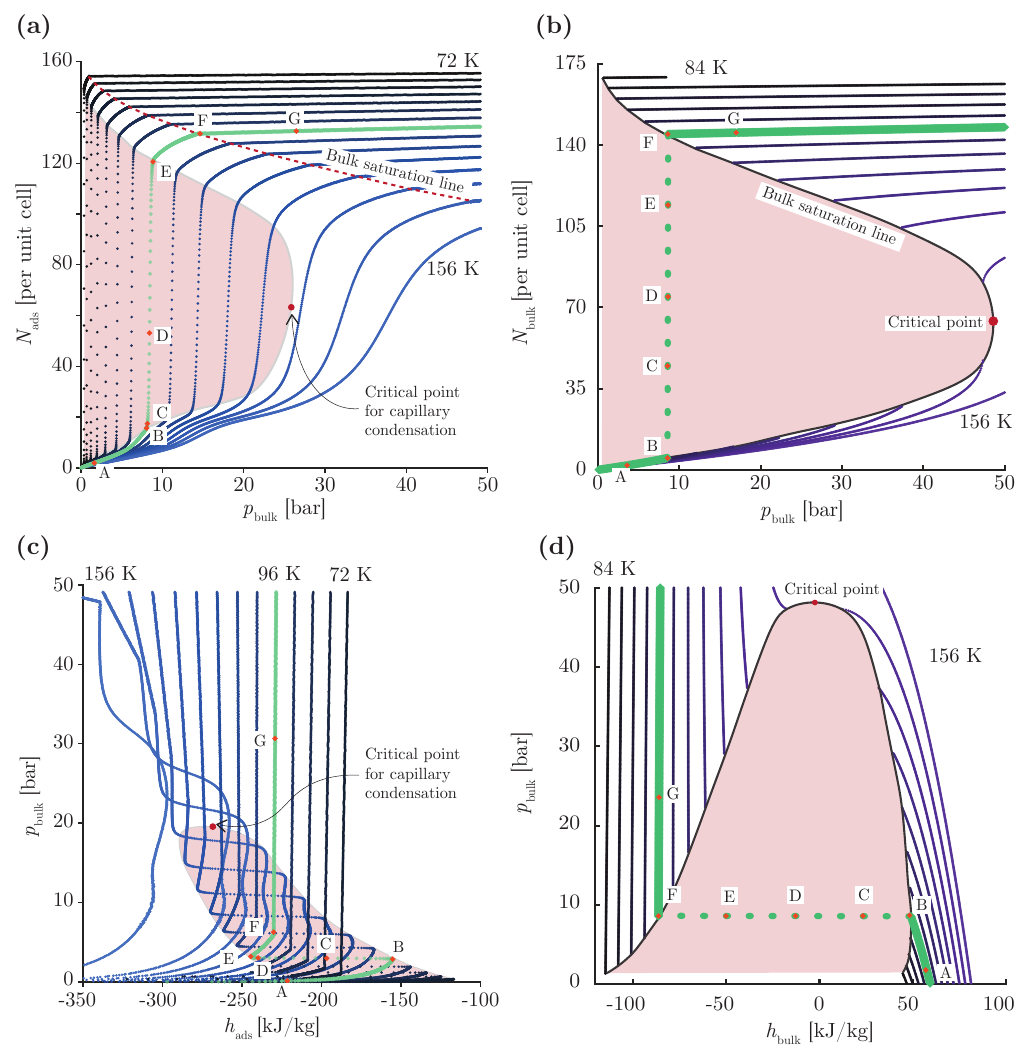}
    \caption{\textbf{Comparison between the confined-fluid and bulk-fluid phase diagrams}\\
    $N-p$ diagram for one unit cell with cell length $a = 24$ \AA{} (a) confined argon and (b) bulk argon. $p-h$ diagram comparison between (c) confined argon and (d) bulk argon.}
    \label{fig:bulk-comparison}
\end{figure}

The bulk phase diagram obtained from the van-der Waals equation is shown here for comparison with the confined-fluid phase diagram. By comparing Fig.~\ref{fig:bulk-comparison}(a) and Fig.~\ref{fig:bulk-comparison}(b), it can be clearly seen that critical point for condensation shifts to a lower value for confined argon(From 49~bar for bulk argon to $\approx$ 25~bar for confined argon). Moreover, the comparison between the confined argon and bulk argon $p-h$ diagram is shown in Fig.~\ref{fig:bulk-comparison}(c) and Fig.~\ref{fig:bulk-comparison}(d).

\bibliography{bibliography}